\documentclass[12pt]{iopart}

\usepackage[utf8x]{inputenc}
\usepackage{iopams}
\usepackage{setstack}
\usepackage{cite}
\usepackage{url}

\usepackage{float}
\usepackage{subfig}
\usepackage{graphicx}
\usepackage{tikz}
\usepackage{sidecap}
\newcommand{\ket}[1]{| #1 \rangle}
\newcommand{\bra}[1]{\langle #1 |}

\graphicspath{{figures/}}

\begin{document}

\title{Noise assisted transport in the Wannier-Stark system}
\author{Stephan Burkhardt$^1$, Matthias Kraft$^1$$^{,2}$, Riccardo Mannella$^3$ and Sandro Wimberger$^1$}

\address{$^1$ Institut f\"ur theoretische Physik, Universit\"at Heidelberg, Philosophenweg 19, 69120 Heidelberg, Germany}
\address{$^{2}$ Blackett Laboratory, Imperial College London, South Kensington Campus, \linebreak SW7 2AZ London, UK}
\address{$^3$ Dipartimento di Fisica ‘E. Fermi’, Universit\`a di Pisa, Largo Pontecorvo 3, 56127 Pisa, Italy}

\begin{abstract}
We investigated how the presence of an additional lattice potential,
driven by a harmonic noise process, changes the transition rate from
the ground band to the first excited band in a Wannier-Stark system.
Alongside numerical simulations, we present two models that capture
the essential features of the dynamics.
The first model uses a noise-driven Landau-Zener approximation and
describes the short time evolution of the full system very well.
The second model assumes that the noise process' correlation time is
much larger than the internal timescale of the system, yet it allows
for good estimates of the observed transition rates and gives a simple
interpretation of the dynamics.
One of the central results is that we obtain a way to control the
interband transitions with the help of the second lattice.
This could readily be realized in state-of-the-art experiments using
either Bose-Einstein condensates or optical pulses in engineered
potentials.


\end{abstract}

\section{Introduction}\label{sc:intro}

During the last years, there have been enormous advances in the
experimental control over Bose-Einstein condensates (BECs) in optical
lattices. States can be prepared with a high degree of control
\cite{Morsch} and the systems can be isolated from external
perturbations for long timescales \cite{PhysRevLett.97.060402}. This
fine control over the system allows to investigate quantum mechanical
effects that were previously very difficult to observe in
experiments. Among these effects are Bloch oscillations, Wannier-Stark
ladders, and Landau-Zener tunneling
\cite{Holthaus1,WilkinsonWSS,GhazalPRA10}. Dynamics similar to the
ones observed in dilute BECs can nowadays also be realized using
purely optical means \cite{Regensburger2012,Szameit07}.

These advances have also paved the way to studying transport properties
of spatially \cite{Billy2008,Jendrzejewski2012,Deissler2010,Modugno,Roati2008}
or temporally disordered systems \cite{ModugnoPrivate}.
Disorder is to some degree present in almost all naturally occurring
systems; in particular it plays a large role in solid state systems.
New effects caused by disorder, such as Anderson localization have
therefore potential ramifications in a wide range of fields.

But systems influenced by noise are not only an interesting subject
due to their omnipresence; a noise process can also allow to fine-tune
the properties of the system, as, for example, observed in the effect of
stochastic resonance \cite{manella1995}. The question naturally arises,
whether a similar effect can be seen in an optical lattice system driven by
a noise process.

While most investigations of noise-driven systems focus on white
\cite{Moss1989} or
exponentially driven noise \cite{Kayanuma2,Pokrovsky,Luchinsky1997,Bray1989,Maier1996}, we use a process known as
harmonic noise in this work. This noise process exhibits a strongly peaked
power spectrum, corresponding to an oscillatory behaviour. It therefore
introduces a characteristic noise frequency and has been observed to
facilitate tunneling processes if this frequency matches the
characteristic frequency of the system \cite{dykman1993,Geier}. In
stochastic dynamics, it is now understood that the escape between
attractors takes place along well defined "paths" in phase space,
which are typically analytic \cite{Luchinsky1997, dykman1992,
  dykman1996, Luchinsky1997PRL, Luchinsky1999}: it is the matching
between these structures and the stochastic noise which facilitates
the escape, a phenomenon observed also in transient dynamics
\cite{Mannella1}.

The system we consider in this work consists of a static tilted
optical lattice which is disturbed by a second, noise-driven
lattice. In the absence of the noisy lattice, the system corresponds to the
well-known Wannier-Stark problem\cite{Avron82}. This Wannier-Stark
system will be used as a reference system throughout this work.

Previous work by Tayebirad \emph{et al.} \cite{GhazalPRA11} reports
results based on numerical simulations of our system. In this paper,
we will investigate the effects of the parameters of the noise process
more thoroughly and introduce two different models that indeed explain
many facets of the behaviour of the noise-driven problem.

In \sref{sc:NWSS} we will introduce the system studied in this paper
and discuss it in the context of non-interacting BECs in optical lattices. We then
introduce two models that allow for an interpretation of the
behaviour of the system in \sref{NWWS:SimplifiedModels}. An
extensive numerical study is then presented in \sref{sc:NumRes};
these results are compared to the models presented in the previous
section and we give an outlook on weakly interacting BECs. We close with a short conclusion of the results in
\sref{sc:concl}.


\section{Noisy Wannier-Stark system}\label{sc:NWSS}
In dimensionless units the Hamiltonian studied in this work reads
\begin{equation}
\hat{H}=- \frac{1}{2} \partial_{x}^2 + V_0\left[ \cos(x) + \cos(\alpha [x - \phi(t)])\right] -  F_0 x  \label{HGhazal}  ,
\end{equation}
where $V_0$ gives the potential depth of the lattice, $F_0$ is the
Stark force, $\alpha$ is a real number\footnote{If this number
  $\alpha$ is irrational, the resulting system is quasi-periodic,
  similar to those used to study Anderson localization by Roati
  \emph{et al.}\cite{Roati2008}. } and $\phi(t)$ is a so-called harmonic
noise process (introduced below). For a conversion between our dimensionless
units, SI-units and common experimental units (applied e.g. in
\cite{GhazalPRA11,GhazalPRA10,ZenesiniPRL}), see appendix
\ref{ssc:units}.

In the absence of the second phase-shifted stochastic lattice, the
Hamiltonian in \eref{HGhazal} corresponds to the well known
Wannier-Stark system \cite{Kolovsky2,Arimondo2011}. A 
quantum state $\ket{\psi}$, which is sufficiently localized in momentum space 
(i.e. much smaller than the Brillouin zone) and prepared in the ground band, 
undergoes Bloch oscillations with the period $T_B=1/F_0$ and partially tunnels into
the first excited band everytime it reaches the edge of the Brillouin zone
(corresponding to an avoided crossing of the respective energy bands). The characteristic time scale and
frequency of the system are thus the Bloch period $T_B$ and Bloch
frequency $\omega_B=2\pi/T_B$ \cite{Kolovsky2}. The
second potential term renders the system stochastic (via the phase
$\phi$) and spatially disordered in the static case
($\phi=\mathrm{const}$) if $\alpha \notin \mathbb{Q}$.

Possible experimental realizations of such systems are for example
BECs in optical lattices \cite{GhazalPRA10, Roati2008}. Similar
Hamiltonians are for example used for precision measurements of forces
acting on the atoms of the BEC\cite{PhysRevA.86.033615}. If the BEC
is either dilute enough and/or interactions are suppressed by tuning with the 
help of a Feshbach resonance, their behavior can be well described by a single particle
Hamiltonian as the one in \eref{HGhazal} (see, e.g., \cite{ZenesiniPRL,GhazalPRA10,Niels2012}). 
In order to better understand how the observed effects would manifest in BECs with
non-vanishing interactions, we will also look at simulations using the
Gross-Pitaevskii equation in \sref{ssc:ChaVar}. There are also recent
experiments using optical pulses in engineered potentials that realize
similar Hamiltonians \cite{Regensburger2012,Szameit07}.

Let us briefly summarize the most important properties of harmonic
noise. Harmonic noise can be imagined as a damped harmonic oscillator
driven by gaussian white noise. It is defined via the two coupled
stochastic differential equations \cite{dykman1993, Geier}
\begin{eqnarray}
&\dot \phi=\mu \\
&\dot \mu=-2\Gamma \mu-\omega_0^2 \phi+\sqrt{4\Gamma T} \xi(t) \label{DefHN}  ,
\end{eqnarray}
where $\Gamma$ is the damping coefficient, $\omega_0$ sets the
characteristic frequency of the noise and $T$ determines the noise
strength. $\xi(t)$ represents gaussian white noise
\cite{mannella2002,Gardiner}. The equilibrium
distributions of the noise variables are of bivariate gaussian type
with first and second moments given by \cite{Geier, mannella2002}
\begin{equation}
 \langle\phi\rangle=0 ,\quad  \langle\mu\rangle=0  ,\quad  \langle\phi^2\rangle=\frac{T}{\omega_0^2} \quad\rm and \quad\langle\mu^2\rangle=T  \label{DefMoments} .
\end{equation}

Just as the damped harmonic oscillator, harmonic noise can be
classified into two different regimes, $\omega_0 > \sqrt{2}\Gamma$ and
$\omega_0 < \sqrt{2}\Gamma$.
In this paper we will only consider the first case, where the noise
shows oscillatory behaviour and its power spectrum has a clear peak at
a frequency $\omega_1=\sqrt{\omega_0^2- 2 \Gamma^2}\approx\omega_0$
for $\omega_0\gg \Gamma$ \cite{Geier, mannella2002, GhazalPRA11}.

While the work of Tayebirad \emph{et al.} provided a first numerical
investigation of this system and showed that the noise properties have
an influence on the observed transport, it could not answer the
question after the mechanism of this influence \cite{GhazalPRA11}.
In order to provide an answer to this problem, we present two simple models that
explain how the noise process acts on the wavefunction in
different regimes.
We then compare the predictions of the two models with a more
systematic numerical study and find that they give a good
agreement with, and provide an intuitive understanding of the relevant dynamics.
Finally, we also look at the stability of the observed effects against
nonlinear interactions as they would, e.g., be present in a realization
using BECs.


\section{Simplified Models}
\label{NWWS:SimplifiedModels}
\subsection{`Noisy' Landau-Zener model}\label{ssc:NLZ}
It turned out that the incommensurability of the two lattices does not
have a great effect on the survival probability \cite{Matthias} due to
the time-dependent phase $\phi(t)$.
We will thus set $\alpha=1$ and present a two state model that allows us to approximate the system's dynamics around
the band edge. The Hamiltonian of \eref{HGhazal} now reads,
\begin{equation}
\hat{H}=- \frac{1}{2} \partial_{x}^2 + V_0\left[ \cos(x) + \cos(x - \phi(t))\right] -  F_0 x  \label{HMat}  .
\end{equation}
The translational invariance (at fixed time) of this Hamiltonian can be recovered by applying the gauge transformation
$\psi= e^{i F_0 x t} \tilde{\psi}$ (changing the frame of reference from the lab system to the accelerated lattice frame) 
\cite{GhazalPRA10,Kolovsky2}. This yields,
\begin{equation}
\hat{H}=  \frac{1}{2}\left(\hat{p}  + F_0 t  \right)^2 +  V_0 \left[ \cos(x) + \cos(x - \phi(t))\right]  \label{Hgauge}  ,
\end{equation}
where we identified $-i \partial_x$ as the momentum operator $\hat{p}$. 
It is instructive to express this Hamiltonian in the momentum basis, this gives
\begin{equation}
\fl \hat{H}=  \int_p dp \frac{1}{2}\left[ \left(p  + F_0 t  \right)^2 \ket{p} \bra{p} + V_0 \left( (1+e^{i \phi}) \ket{p} \bra{1+p} + (1+e^{- i \phi})\ket{1+p} \bra{p} \right)\right]  \label{HDecom},
\end{equation}
from which it is clear that it only couples states with $\Delta p= p - p'  \in \mathbb{Z}_0$ (due to the time evolution).
The momentum states can thus be written as
$\ket{p}=\ket{n+k}$, with $n \in \mathbb{Z}_0$ and $k \in [-0.5, 0.5) \subset \mathbb{R}$ being the quasimomentum of the system. 
Due to the conservation of quasimomentum, the Hamiltonian in \eref{HDecom}
can be decomposed into independent Hamiltonians each acting
on a subsystem of constant quasimomentum $k$ \cite{GhazalPRA10, Niels}. A system initially in state $\ket{p_0}=\ket{k_0 + n}$ thus evolves according to the
(tridiagonal) matrix Hamiltonian,
\begin{equation}
        \fl \hat{H}_{k_0} =\frac{1}{2}\left(
         \begin{array}{ccccc}
  		\ddots &V_0 (1+e^{i \phi})  &  &  & \mbox{\Huge $\mathbf 0$}\\
 			  & \color{blue}{\bf{(k_0 - 1 + F_0 t)^2}} &\color{blue}{\bf{V_0 (1+e^{i \phi}) }}  &  &  \\
  			  & \color{blue}{\bf{V_0(1+e^{-i \phi}) }}  & \color{blue}{\bf{(k_0 +F_0 t)^2}} & V_0(1+e^{i \phi})  & \\
  		 	&  & V_0(1+e^{-i \phi})  & (k_0 + 1 +F_0 t)^2 & \\
  		 \mbox{\Huge $\mathbf 0$} &  &  & V_0(1+e^{-i \phi}) &  \ddots \\
         \end{array}\right) ,
 \end{equation} 
where the $V_0 (1+e^{\pm i \phi})$ terms stem from the optical lattices and $\phi$ enters as a complex phase.
The dynamics of the full system around an avoided crossing of the ground and first
excited energy band (at the edge of the Brillouin zone) can be approximated by the \textbf{\color{blue}{highlighted}} part of the
above matrix\footnote{This corresponds to reducing the system to the
  two lowest energy states. At the edge of the Brillouin zone, this
  approximation is very accurate as long as the off-diagonal coupling
  terms are not too large, since only the lowest energy states
  contribute to the first excited as well as the ground band.} with $k_0=0.5$, 
i.e. the value at the edge of the Brillouin zone\cite{GhazalPRA10, Niels}. This yields a reduced Hamiltonian
 \begin{equation}
 \hat{H}_{N, LZ} =\frac{1}{2}\left(
         \begin{array}{cc}
  		  - F_0 t &  V_0 (1+e^{i \phi})\\
  					V_0 (1+e^{-i \phi}) &  F_0 t \\
         \end{array}\right) \label{HLZNoise} ,
\end{equation}
where we additionally subtracted $\frac{1}{4} + (F_0 t)^2$, i.e. we shifted the absolute value of the energy scale.
This 2-by-2 matrix can be seen as an effective `noisy' Landau-Zener (LZ) Hamiltonian; its \emph{instantaneous}
eigenvalues are given by,
\begin{equation}
E_{N;1,2}= \mp \frac{1}{2} \sqrt{(F_0 t)^2 +2 V_0^2(\cos(\phi(t))+1)} .
\end{equation}
We can thus introduce an \emph{effective band gap} as a function of time by averaging over the probability distribution of the noise process $\phi$.
At $t=0$, this leads to
\begin{equation}
\mbox{\small $\Delta$} E_{\rm eff}=V_0 \left[ \langle \sqrt{2(\cos(\phi(0))+1)}\rangle \pm \rm{Std}\left(\sqrt{2(\cos(\phi(0))+1)})\right) \right] \label{EffBG} ,
\end{equation}
where $\langle f(\phi) \rangle$ denotes the average over the noise process and `Std' means standard deviation (see \fref{fig:EVLZNoise} for a schematic
representation of $\Delta E_{\rm eff}$).

In the case of a single (noiseless) optical lattice only the constant term in the anti-diagonal of the Hamiltonian in \eref{HLZNoise} is present and
one recovers the standard LZ model \cite{Zener, Landau, Stuckel, Majo}. Starting in the ground state at time $t=-\infty$, the probability
to remain in that state until time $t=\infty$ is given by the Landau-Zener formula \cite{Zener, Landau, Stuckel, Majo},
\begin{equation}
 P_{sur}(t=\infty)=1-\exp\left(-\frac{\pi V_0^2}{2 F_0}\right) \label{LZFormula} .
\end{equation}
This allows to give estimates for the survival probability in the `noisy' LZ model by replacing $V_0^2$ with $\Delta E_{\mathrm{eff}}^2$ in \eref{LZFormula}.

The two state approximation around an avoided crossing is only
applicable if the initial momentum distribution is sufficiently localized in
momentum space (i.e. much smaller than the Brillouin zone) \cite{ZenesiniPRL} and if the characteristic time of
the system is larger than the transition time in the two state model
\cite{Vitanov1, Vitanov2}, i.e. the transition time should be smaller than
one \emph{Bloch period}.


\subsection{Quasistatic model}\label{ssc:StToy}

Since the noisy LZ model is limited to short timescales and can only
be applied if both lattices in \eref{HGhazal} have the same
wavelength, we will present another model which is able to describe
the behaviour of the system in cases where those
conditions are not met.

This model is based on a quasistatic approximation of the noise
process; $\phi(t)$ is replaced by a term $\beta t$ linear in time. The
resulting system is analyzed and its observables are averaged
according to the properties of the noise process. In the following
paragraphs, we will provide a more detailed description of this model.

For short timescales, the dynamics of the system can be approximated
by a Taylor expansion of the noise term $\phi(t)$:
\begin{eqnarray}
  \label{eq:tm:1}
  \phi(t + \epsilon) &= \phi(t) + \epsilon \frac{\rmd}{\rmd t} \phi(t)  +
  \mathrm{O}(\epsilon^2) =\phi(t) + \epsilon \mu + \mathrm{O}(\epsilon^2).
\end{eqnarray}
A rigorous mathematical treatment of the $\epsilon ^2$ term in
\eref{eq:tm:1} is far from trivial\footnote{As the second time
  derivative of the noise term $\phi(t)$ is not well-defined, it is
  necessary to look at a value $\langle \phi(t) \rangle$ which is
  averaged over a timescale, which is much shorter than the characteristic one of the
  system.}. Nevertheless, it is clear that it depends on
$\frac{\rmd^2}{\rmd t^2}
\phi(t) = \frac{\rmd}{\rmd t} \mu(t)$ as given in \eref{DefHN} and is thus
negligible for $\omega_0,\Gamma \ll 1$.  In this case, \eref{HGhazal}
can therefore be approximated as
\begin{equation}
  \label{eq:tm:2}
  H = - \frac{1}{2} \partial_x ^2 + V_0 \cos(x) + V_0 \cos\big(\alpha(x-\phi_0-\beta t)\big) + Fx,
\end{equation}
where the noise variable $\mu(t)$ has been replaced by a constant term
$\beta$. \Eref{eq:tm:2} describes a tilted bichromatic optical lattice
system where the two lattices have a constant relative velocity
$\beta$.
\begin{figure}[t]
  \centering
  \input{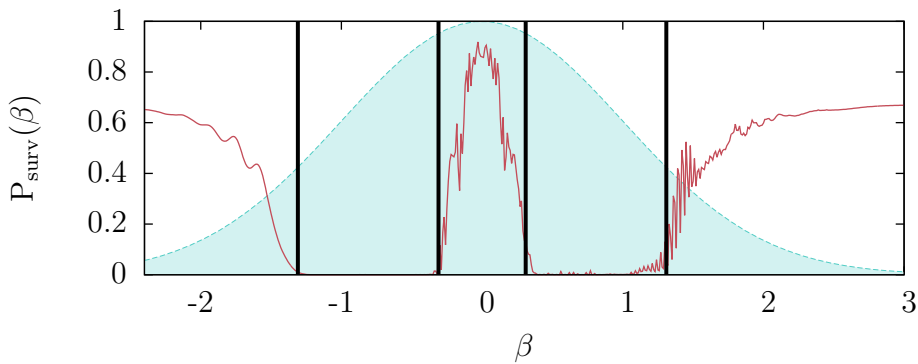}
  \caption{The fluctuating red curve shows the survival probability for the
    noiseless bichromatic lattice for a relative lattice velocity of
    $\beta$. The parameters are $\alpha=0.61, F_0=0.00762,
    \Gamma=0.00762$ and $V_0=0.125$. The black bars mark the values of
    $\beta$ which lead to an intertwined position of the barriers (see
    \fref{fig:tm:tunnel1}). The blue gaussian curve of variance $T$
    gives the probability distribution of $\mu$ given in
    \eref{DefMoments}. In the quasistatic model, we set $\beta=\mu$
    and $P_{\rm sur}(\beta)$ is therefore averaged over this gaussian
    distribution.}
  \label{fig:tm:betagraph}
\end{figure}
The dynamics of such a system can be seen in \fref{fig:tm:betagraph}.
The fluctuating red curve shows the survival probability of the state
in the ground band after $t= T_B$. It can be seen that this survival
probability drops sharply for intermediate values of $\beta$
($|\beta|\approx 0.31-1.31$, region between the black vertical
lines). This corresponds to strongly enhanced interband transport for
these parameters.

\begin{figure}[b]
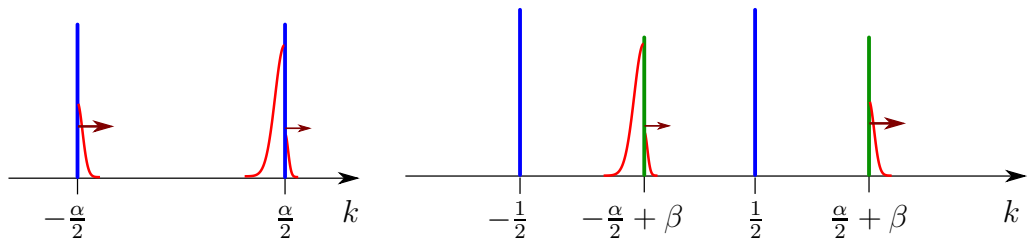

  \centering
  \input{figures/figure2a.tex}
  \input{figures/figure2b.tex}
  \caption{Left: Effect of a single lattice on
    momentum eigenstates. Right: Combined effect of a static and moving lattice.}
  \label{fig:tm:tunnel1}
\end{figure}

As in \sref{ssc:NLZ}, the
key to understanding lies in considering the effect of optical
lattices on the wavefunction in momentum space. Let us first look at a single tilted
optical lattice $V_0 \cos(\alpha x)$. We assume that the timescale of the
Landau-Zener transition from the ground band to the first excited band
is short compared to a Bloch period and is thus accurately described
by the Landau-Zener model.  Due to the static force F, a
momentum-eigenstate scans through momentum space; once it reaches a
momentum $p=\pm\frac{\alpha}{2}$, it undergoes a reflection on the
lattice with the probability of $P_{\mathrm{sur}}$ given in
\eref{LZFormula}. This process is schematically displayed in
\fref{fig:tm:tunnel1} (left).
For the ground band, the effect of the lattice can therefore be
represented as a pair of barriers at $\pm \frac{1}{2}\alpha$ (blue
barriers in left of \fref{fig:tm:tunnel1}), which
trap the state in between them. This momentum range between the barriers
corresponds to the ground band. For the lattice moving with velocity
$\beta$, the position of these barriers in momentum space is shifted
to ($\pm \frac{\alpha}{2} + \beta$). Adding up the effects of both,
static and moving lattice thus results in the presence of two pairs of
barriers as seen in the right part of \fref{fig:tm:tunnel1}. If the
two pairs of barriers are in an intertwined position (as shown in the
right of \fref{fig:tm:tunnel1}), reflections on
the moving lattice can help the state escape from the ground
band. This can be understood from the right part of
\fref{fig:tm:tunnel1}. If the state `hits' the barrier at
$k=-\frac{\alpha}{2} + \beta$ from the left (green barriers in \fref{fig:tm:tunnel1}), it can either be
transmitted or reflected. While the transmitted fraction of the state is
still trapped in-between the barriers at $k=\pm\frac{\alpha}{2}$, the
reflected fraction appears at the right side of the barrier at
$k=\frac{\alpha}{2} + \beta$ and has thus escaped the ground band of
the non-moving lattice. This can be interpreted as absorbing momentum
from the moving lattice by means of a Bragg reflection and thus being
`boosted' out of the ground band. It
should be noted that the same process can also take place for negative
beta (transporting the state to the left of ground band in figure
\ref{fig:tm:tunnel1}, therefore also out of the ground band). Transport to higher energy bands is therefore
enhanced for intertwined barriers. This is supported by the data
visible in \fref{fig:tm:betagraph}, where the values of $\beta$ which lead to
an intertwined configuration are the area in-between the black vertical
lines. These values of $\beta$ do indeed correspond to a heavily suppressed
survival probability in the ground band and thus to strongly enhanced transport.

Keeping this picture in mind, we can understand how the noise process
acts upon the momentum states of the system. The constant $\beta$ is
responsible for the relative position of the barriers in momentum
space.  In the real system, $\beta$ is given by the noise variable
$\mu(t)$ and is thus a function of the time. The positions of the
barriers of the noise driven lattice thus move in momentum space.

In our quasistatic model we assume $\mu(t)$ to be a constant $\beta$,
which is a realistic approximation as long as the noise process is
slow, i.e. $\mu(t)$ changes slowly with time.  For $\mu(t)=\beta$ the
survival probability in the ground band is given by
\fref{fig:tm:betagraph}. To obtain an approximation for the behavior
of the full, noise-driven system, we thus average this survival
probability $P(\beta)$ over the equilibrium distribution of $\mu(t)$,
which is a gaussian of width $T$ (see \eref{DefMoments}) as pictured
in \fref{fig:tm:betagraph}.

While we would expect this model to be accurate for slow noise
processes, it ignores the influence the non-static nature of $\mu(t)$
has on the Landau-Zener transitions and does also not
account for time-correlations of the noise-process. Especially for a
fast noise process where $\mu(t)$ changes significantly during one
Bloch period, we therefore expect discrepancies between the
quasistatic model and the full system \cite{Stephan}.

\section{Numerical Results}\label{sc:NumRes}

We compare now the predictions made by the two models
introduced in the previous section with data from the full system described 
by the Hamiltonian in \eref{HGhazal}.

The most important property of the noise process $\phi$ is its
\emph{variance}. In the following, we will analyze the survival
probability of the system in the ground band for the two cases: $\rm
Var(\phi)=constant$ (section \ref{ssc:ConVar}) and $\rm
Var(\phi)=changing$ (section \ref{ssc:ChaVar}). The latter is realized
by keeping $\omega_0$ constant and varying $T$.

\begin{SCfigure}[][!ht]
  \input{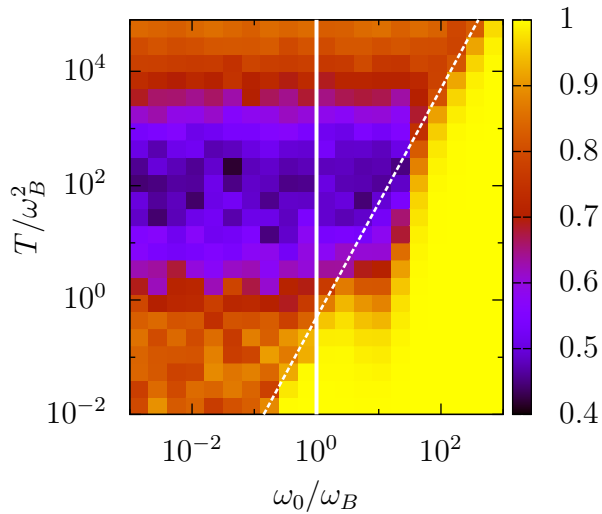}
  \caption{Survival probability of the system described by
    \eref{HGhazal} in the ground band (colour coded) after $t= T_B$,
    i.e. after one transition, versus the rescaled noise frequency
    $\omega_0/\omega_B$ and the noise strength $T/\omega_B^2$. The
    vertical solid line represents a line of constant $\omega_0$. The
    inclined dashed line represents a line of constant variance.  Data
    from numerical simulations of the Hamiltonian in \eref{HMat}, for
    details see appendix \ref{ssc:nummeth}. The parameters are
    $F_0=0.00762$, $\Gamma=0.00762$ and
    $V_0=0.125$.}\label{fig:2dplot}
\vspace{-20pt}
\end{SCfigure}

\Fref{fig:2dplot} shows the survival probability in the ground
band after one Bloch period. The survival probability strongly depends
on the noise parameters $T$ and $\omega_0$.

\subsection{Constant variance, i.e. $T/\omega_0^2=\rm const$ .}\label{ssc:ConVar}

\begin{figure}[t]
  \centering
\input{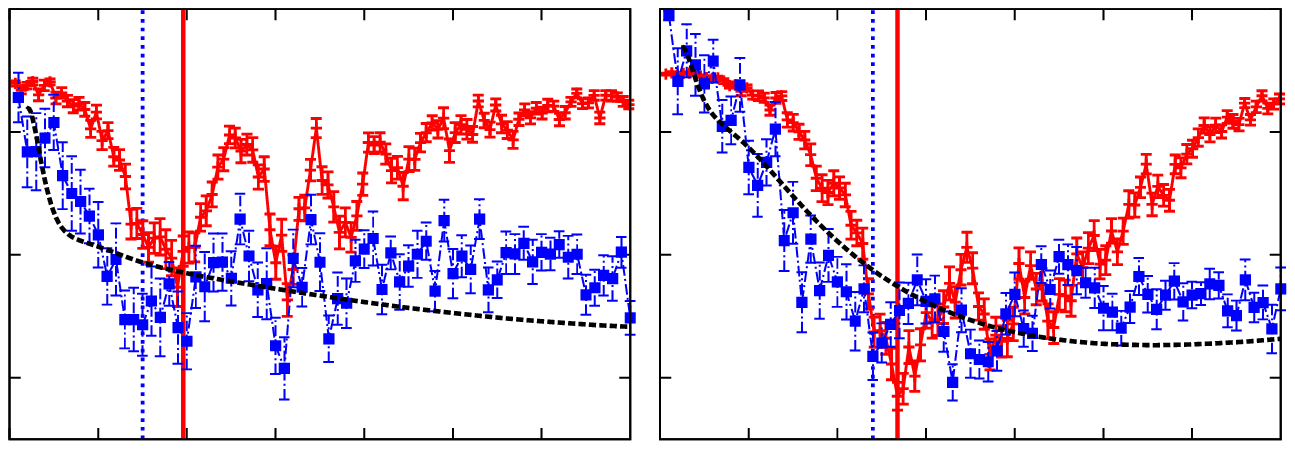} 
\vspace{-10pt}
\caption{Survival probability in the ground band at $t=T_B$ versus the
  rescaled noise frequency $\omega_0/\omega_B$.  Shown are data for
  the real system (blue dashed lines), the `noisy' LZ model (red solid
  line) and the quasistatic model (black dotted line). The vertical lines
  indicate the position of the first minimum in the survival
  probability for their respective data sets. The parameters are
  $F_0=0.00762$, $\Gamma=0.00762$, $\langle\phi^2\rangle=0.5$ and
  $V_0=0.0625$ (left) and $V_0=0.125$ (right).}
\label{fig:NumResults5}
\end{figure}

Here, we show the survival probability in the ground band after one
Bloch period for constant variance (cut along the dashed line in
\fref{fig:2dplot}) versus the noise frequency. We focus on the
region of the initial decrease and minimum of the survival
probability, i.e. $\omega_0/\omega_B \lesssim 14$.  
\Fref{fig:NumResults5} compares numerical data from the `noisy' LZ
model in \eref{HLZNoise}, with data from the full Hamiltonian in
\eref{HMat} for two values of the potential depth $V_0$.

There is an excellent qualitative and a good quantitative agreement
between the `noisy' LZ predictions and the simulations of
the full system. The `noisy' LZ accurately reproduces the
initial decay of the survival probability (as obtained for the stochastic Wannier-Stark system) for increasing noise
frequency and gives a good approximation for the position of the
minimum. Yet, it systematically overestimates the survival
probability. This overestimation is less pronounced for higher
potential depths $V_0$ (compare  \fref{fig:NumResults5} (left)
and \fref{fig:NumResults5} (right)). 
Most importantly, the position of the minimum changes with varying
potential depth $V_0$.
The quasi-static model also shows a decrease in the survival probability with increasing frequency, but cannot reproduce any of the finer
features visible in the stochastic Wannier-Stark system and the `noisy' LZ-model.

The varying position of the minimum leads us to the next figure, \fref{fig:NOmegaMin}, in which the
position of the first minimum in the survival probability is plotted
versus the potential depth $V_0$. Here, the predictions of the `noisy'
LZ model and the quasistatic model are compared to the full system;
also the effective band gap $\Delta E_{\rm eff}$ is plotted as a
function of $V_0$.  The data for the full system shows that there
exists a linear relationship between the position of the minimum in
\eref{fig:NumResults5} and the potential depth $V_0$. This
relationship is accurately reproduced by the `noisy' LZ
model, but it can not be seen in the quasistatic model.
The fact that the quasistatic model can not account for the position of the
minimum is easily understood by looking at \fref{fig:NumResults5}.
Since the model does not predict the local minima visible in
$P_\mathrm{sur}$ of both the full system as well as the noisy LZ
model, it is clear that the position of the first minimum will in
general be at much larger values of $\omega_0$.

Another observation that can be made in \fref{fig:NOmegaMin} is that
the slope of the linear curve (for the full system and the LZ model)
is approximately given by the one for the effective band gap. In fact,
linear fits to the data give a slope of $1.93$ for the real system and
$1.76$ for the LZ model, compared to $1.88\pm0.15$ for the effective
band gap \eref{EffBG}. Hence, within errors, the position
of the minimum is given by the effective band gap plus a constant
off-set. The linear relationship also holds for different values of
$F_0$ \cite{Matthias}.

\begin{SCfigure}[][h!]
   \input{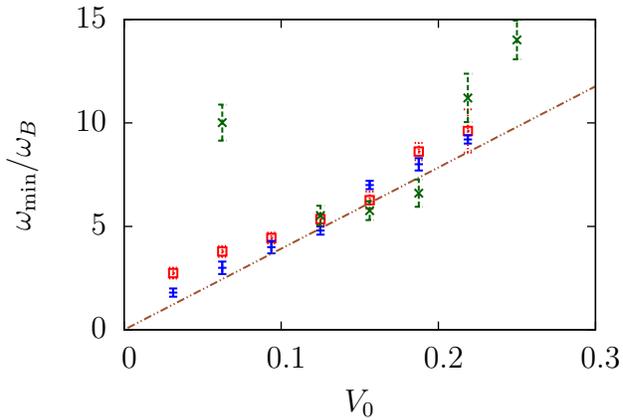}
   \caption{Frequency position of minimum versus potential depth
     $V_0$. Shown are the data for the real system (blue lines), the
     `noisy' LZ prediction (red squares), the quasistatic model
     prediction (green crosses) and the effective band gap $\Delta
     E_{\rm eff}$ (dashed-dotted line). The parameters for the
     numerical simulations are $F_0=0.00762$, $\Gamma=0.00762$,
     $\langle\phi^2\rangle=0.5$.}
\label{fig:NOmegaMin}
\end{SCfigure} 

\begin{SCfigure}[1.0][h!]
\centering
  \input{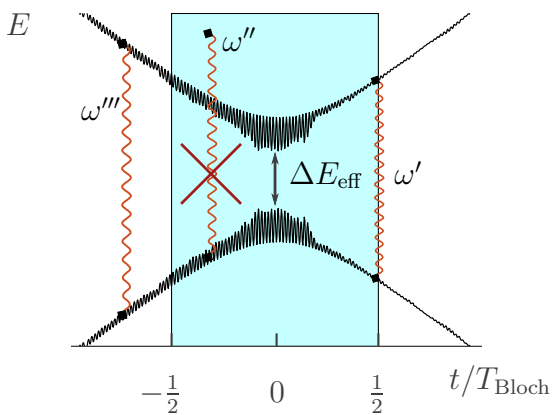}
  \caption{Instantaneous eigenvalues (solid black lines) of the
    `noisy' LZ model for a typical noise realization. Schematically
    represented are $\Delta E_{\rm eff}$ and transitions between
    energy states (red wiggly lines). The blue shaded area extends over one Bloch period and approximates the energy band structure of the `noisy' WSS around an avoided crossing.
    The crossed out transition is not allowed.}
  \label{fig:EVLZNoise}
\end{SCfigure}

The linear dependence of the minimum on $V_0$ can be interpreted as
follows. As mentioned in section \ref{sc:NWSS}, the harmonic noise
`feeds' an energy of $\approx \omega_0$ into the system. Once this
energy is high enough to overcome the band gap between the ground and
first excited band, given by $\approx \Delta E_{\rm eff}$, the system
can be excited into the upper band via a `phononic' excitation
process\footnote{`Phononic' is not to be taken literally, but
  indicates that the energy is transferred to the atoms via an
 optical lattice shaking/vibrating with a more or less well-defined
 frequency of $\approx \omega_0$.} (red wiggly lines in \fref{fig:EVLZNoise}).

The overestimation of the survival probability by the `noisy' LZ model
occurs because the `LZ' model overestimates the separation of the
bands in the real system far away from an avoided crossing (the model only works in the blue shaded region of figure \ref{fig:EVLZNoise}); hence
transitions to higher bands are suppressed (the effective bandgap increases linearly with time $t$ for large times, see \fref{fig:EVLZNoise}). For large $V_0$, only
transitions close to an avoided crossing can occur, but for smaller
$V_0$ transitions across the full Brillouin zone happen in the real
system and this can not be ignored.

The early rise of the survival probability in the `noisy' LZ model is
due to its two band nature. If the energy fed into the system by the
noise is too large, there exists no target state for the transition (see crossed out transition in 
\fref{fig:EVLZNoise}). In the real Wannier-Stark system, this rise
happens only at higher $\omega_0$ because it is not just a two band
system and transitions to higher bands are actually possible.

\subsection{Changing variance, i.e. constant $\omega_0$ and varying $T$.} \label{ssc:ChaVar}
\begin{figure}[b]
  \centering
\input{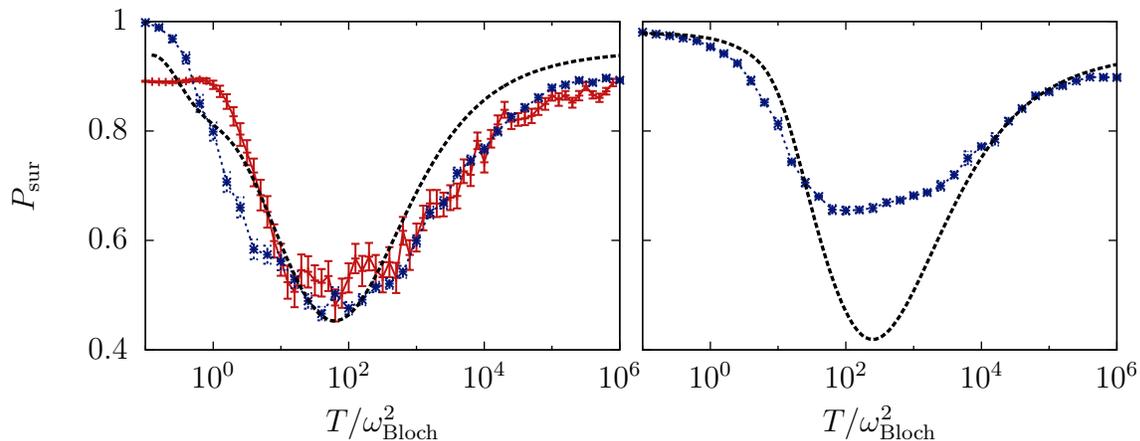}
\caption{Survival probability in the ground band for fixed noise
  parameter $\omega_0$. Plotted are the results from the full system
  (blue stars), the predictions by the noisy LZ model (red line) as
  well as the predictions from the quasistatic model (black dashed
  line). Parameters are $F_0=0.00762, \Gamma=0.00762, V_0=0.125$ and
  $\alpha = 1$ (left) and $\alpha=0.618$. The plotted quantity is the
  survival probability after one Bloch period
  $P_\mathrm{sur}(t=T_\mathrm{Bloch})$ (left) and the average survival
  rate for long timescales
  $P_\mathrm{sur}(t+T_\mathrm{Bloch})/P_\mathrm{sur}(t)$
  (right).}
  \label{fig:compareV2}
\end{figure}

In \fref{fig:2dplot} we observe the following: once $\omega_0$ assumes a small
enough value it stops having a discernible influence on the survival
probability $P_\mathrm{sur}$ in the ground band. In this section we
will therefore take a more detailed look at how $P_\mathrm{sur}$ in
this regime depends on $T$ while keeping $\omega_0=1$ constant.

On the left side of \fref{fig:compareV2}, we see the survival
probability of a system initially prepared in the ground state in
comparison to the predictions from the two models. The observable
$P_\mathrm{sur}(t=T_\mathrm{Bloch})$ is the same as previously looked at in
\fref{fig:NumResults5}. Good qualitative agreement between the full
system and both models is observed. Furthermore, within the
numerical errors, the minimum of the survival probability is correctly
predicted by both models. There are, however, quantitative discrepancies
for small (noisy LZ model) as well as large values of $T$
(quasistatic model).

On the right side of \fref{fig:compareV2}, a system with two different
lattice constants is studied ($\alpha = 0.618$). The quantity analyzed
in this case is the large time survival probability of the ground state
per Bloch period
$P_\mathrm{sur}(t+T_\mathrm{Bloch})/P_\mathrm{sur}(t)$ for $t>5$. This
quantity is similar to the previously examined survival probability
$P_{\mathrm{sur}}(t=T_\mathrm{Bloch})$, but it is derived from the long
term exponential decay of the initial state. No results for the noisy
LZ model are shown since this model cannot make any predictions for
the case $\alpha \neq 1$. For the quasistatic model, good quantitative
agreement with the full system is observed for most values of
$T$. Nevertheless, significant differences are visible around the
predicted minimum. These differences, however, result from the
limitations of the quasistatic model and vanish if $\omega_0$ as well
as $\Gamma$ are chosen small enough. In this case, where $\alpha \neq
1$, the quasistatic model is very successful in quantitatively
predicting the long-term survival probabilities, especially if
$\Gamma$ as well as $\omega_0$ are small \cite{Stephan}.

\begin{SCfigure}[1.0][h]
  \input{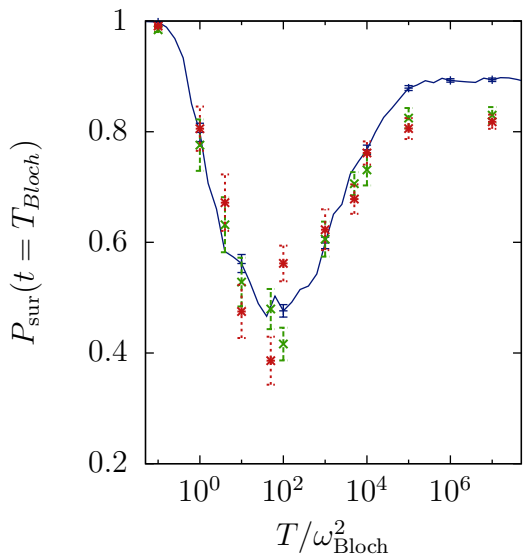}
  \caption{Survival probability in the ground band after one Bloch
    period using the Gross-Pitaevskii equation (same parameters as
    left of \fref{fig:compareV2}). Results from the (linear)
    Schrödinger equation are shown in solid blue, while green crosses
    show results for $N=5\cdot10^4$ and red stars for $N=10^5$
    atoms in a cigar-shaped condensate. The simulated setup was the one used in the experiments
    reported in \cite{Sias07} and is described in
    \ref{ssc:nummeth}. The used atom numbers correspond to dimensionless
    nonlinearity parameters of $C\approx0.08$ (green crosses) and
    $C\approx0.13$ (blue stars), c.f. ref. \cite{Sias07,Arimondo2011}.}
  \label{fig:num:nonlinear}
\end{SCfigure}

In order to understand whether the described effects would be visible
in a realistic experimental setup using BECs, it is important to know
how they are affected by interactions between the atoms in a
condensate. Using the 3D Gross-Pitaevskii equation, we performed
numerical calculations for an experimental setup similar to the one
used in \cite{Sias07}: ${}^{87}$Rb atoms that are placed in a
cigar-shaped optical dipole trap with $250 \mathrm{Hz}$ radial and $20
\mathrm{Hz}$ longitudinal frequency. The results can be seen in figure
\ref{fig:num:nonlinear} and show that, while clear differences in the
survival probability exist, the enhanced tunneling rate for
intermediate noise frequencies persists and is hardly influenced by
the nonlinearity. It should especially be noted that for a
nonlinearity parameter (as defined in \cite{Sias07}) up to
$C\approx0.13$, the position of the minimum is hardly changed from the
single particle dynamics \footnote{Nonlinearity parameters are only
  given approximately, as they depend on the maximal density of the
  condensate. Since our setup is governed by a stochastic equation,
  the peak density of the condensate is not only time-dependent due to
  the interband tunneling \cite{PhysRevA.72.063610}, but also
  fluctuates for different noise realizations.}.


\section{Conclusion}\label{sc:concl}

We have investigated the transport properties of a
bichromatic Wannier Stark problem under the influence of harmonic
noise. Our main interest was understanding the circumstances under
which the noise process can enhance or suppress the transport in
momentum space.

An extensive numerical study of the influence of the noise
parameters revealed that the system's behaviour can be
divided into two different regimes, as seen in \fref{fig:2dplot}. 
In the regime of low variance of the noise process, the noise only
has very little influence, while in the regime of large
variance, the properties of the system almost exclusively depend on
the noise strength $T$. In this regime of large variance, the
transport in momentum space has a clear maximum for a certain value of $T$.

We presented two different models that both offer an interpretation of
these results. The noisy LZ model approaches the system as a
noise-driven Landau-Zener transition and argues that transport should
be maximized if the noise frequency matches the band gap between the
ground and the first excited band. While this model can only predict
results for the case where both lattices have the same wavelength, it
reproduces the numerical results of the full system in this case very
well. Furthermore, it gives an accurate prediction of the noise
parameters necessary to maximize transport (see
\fref{fig:NOmegaMin}). The quasistatic model assumes the noise process to be slow compared to
the timescales of the noise-free system.
Linking the relative velocity of the optical lattices to their action
upon the wavefunction in momentum space, this model predicts that the transport rate only
depends on the noise parameter $T$ and achieves a maximum for
intermediate values of $T$.
Within the low-variance regime seen in \fref{fig:2dplot}, this
prediction matches very well with the results of the full system, even
if full quantitative agreement is only reached for very slow noise
processes. Our quasistatic model, where the particle is effectively kicked
by the moving second lattice, see \fref{fig:tm:tunnel1}, may turn out to be
relevant as well for the understanding of damping effects of Bloch oscillations
in `noisy' solid-state devices \cite{Dekorsy2000,Leo1998}.

Since experiments that study the expansion of the wavefunction in
noise-driven lattices are underway \cite{ModugnoPrivate}, an
experimental realization of our system and a comparison between
experimental and numerical data would be interesting. As shown at the
end of \sref{ssc:ChaVar}, the observed effects are fairly stable
against interactions between the atoms of a BEC and should therefore
be easily observable.
Besides possible
realizations with BECs in optical lattices \cite{Roati2008}, purely
optical techniques such as the one of \cite{Regensburger2012} could
also be used. While control over the BEC using a noise process is subject to large
fluctuations, a deterministic bichromatic lattice, as presented in the
beginning of \sref{ssc:StToy}, could be used to control the transport
between different bands with a high degree of precision.
In \fref{fig:tm:betagraph} we, e.g., can see that with small changes
of the relative lattice velocity $\beta$, the tunneling rates can be
easily tuned by various orders of magnitude.


\ack

We thank Niels Lörch for help and discussions at an early stage of
this work. Furthermore, we are grateful for the support of the
Excellence Initiative through the Heidelberg Graduate School of
Fundamental Physics (Grant No. GSC 129/1), the DFG FOR760, the
Heidelberg Center for Quantum Dynamics and the Alliance Program of the 
Helmholtz Association (HA216/EMMI).


\appendix
\label{sc:appen}
\section*{Appendix}
\section{Numerical methods}\label{ssc:nummeth}

The data for the `noisy' LZ model is obtained by numerically integrating the Schr\"odinger equation corresponding to Hamiltonian \eref{HLZNoise}.
The survival probability has been obtained by starting in the
\emph{diabatic} ground state (i.e., in the momentum eigenbasis of the uncoupled problem
  \cite{Niels2012}) at $t=-0.5 T_B$ and then propagate
it to $t=0.5 T_B$. The survival probability at each time step is obtained by projecting onto the \emph{diabatic} first excited state \cite{GhazalPRA10, Matthias}.
Then the survival probability is averaged over the last 10\% of the total integration time to estimate the asymptotic value of the survival 
probability. In the end, we average over 100 noise realisations and
plot the standard deviation of the average survival probability as
error bars.

The data for the noise-driven Wannier-Stark problem are generated in a
similar way, by integrating the time-dependent Schr\"odinger equation
for a 1D or 3D wavefunction. At $t=0$, the system is prepared in the ground state of
the bichromatic lattice potential superposed by a small harmonic trap
potential (to keep the size of the initial wavefunction finite). At
$t=0$, the harmonic trap potential is disabled and replaced by the
static force $F$. The survival probability in the ground band is
measured by integration over the momentum distribution using
appropriate boundaries (see
\cite{Stephan,TayebiradThesis,GhazalPRA10,Niels2012} for more
details). While simulations for single-particle wave function were
performed in one dimension, realistic results for the Gross-Pitaevskii
equation can only be achieved using simulations in three dimensions
(for the used algorithm, see \cite{Stephan}). For realistic values of the
nonlinearity parameter, we used the values of the setup described in
\cite{Sias07}: ${}^{87}$Rb atoms in an optical lattice
($d_l=426\mathrm{nm}$) that are initially loaded into a cigar-shaped, quasi-one-dimensional
optical trap, with $250\mathrm{Hz}$ radial and $20\mathrm{Hz}$ longitudinal confinement.

\section{Unit conversion}\label{ssc:units}
\begin{table}[H]
 \centering
\footnotesize
\caption{Summary of the new unit system. The table gives relations between our set of dimensionless units, SI-units and a set of common experimental units
\cite{GhazalPRA10, GhazalPRA11, ZenesiniPRL}.
Here, $M$ is the mass of the atoms in the BEC and $k_L$ is the wavelength of the laser light used to generate
the optical lattice. $E_{\rm rec}$ sets the characteristic energy scale of the system. }
    \begin{tabular}{ | l | l | l | l |l|}  
\hline
     & Energy & Momentum &\\ \hline
    1 photon exchange & $E_{\rm rec}= \frac{\hbar^2 k_L^2}{2 M}$ & $p_{\rm rec}= \hbar k_L$ &\\ \hline
    \hline
     & SI-units & dimensionless & experimental \\ \hline
    Energy & $E_{\rm SI}$ & $E= \frac{E_{\rm SI}}{8 E_{\rm rec}}$ & $E_{\rm exp}= \frac{E_{\rm SI}}{E_{\rm rec}}$ \\ \hline
    Time & $t_{\rm SI}$ & $t= t_{\rm SI} \frac{8 E_{\rm rec}}{\hbar}$ & $t_{\rm exp}= t_{\rm SI}$ \\ \hline 
    Space & $x_{\rm SI}$ & $x=2 x_{\rm SI} k_L$  &$x_{\rm exp}=2 x_{\rm SI} k_L$ \\ \hline
    Force& $F_{\rm SI}$ & $F_0= \frac{F_{\rm SI}}{16 E_{\rm rec} k_L}$ & $F_{\rm exp}= \frac{F_{\rm SI}\pi}{E_{\rm rec} k_L}$\\ \hline
    Potential & $V_{\rm SI}$ & $V_0= \frac{V_{\rm SI}}{8 E_{\rm rec}}$  &$V_{\rm exp}= \frac{V_{\rm SI}}{E_{\rm rec}}$\\ \hline
    \end{tabular}
\end{table}


\section*{References}

\bibliography{bibliography}
\end{document}